\documentclass[pdflatex,sn-mathphys-num]{sn-jnl}

\usepackage{threeparttable} %
\usepackage{adjustbox}  
\usepackage{tabularx} %
\usepackage{graphicx}%
\usepackage{multirow}%
\usepackage{amsmath,amssymb,amsfonts}%
\usepackage{amsthm}%
\usepackage{mathrsfs}%
\usepackage[title]{appendix}%
\usepackage{xcolor}%
\usepackage{textcomp}%
\usepackage{manyfoot}%
\usepackage{booktabs}%
\usepackage{algorithm}%
\usepackage{algorithmicx}%
\usepackage{algpseudocode}%
\usepackage{listings}%

\begin{document}

\title[Article Title]{Pseudo Channel: Time Embedding for Motor Imagery Decoding }

\author*[1,2]{\fnm{Zhengqing} \sur{Miao}}\email{mzq@tju.edu.cn}

\author[1]{\fnm{Meirong} \sur{Zhao}} 

\affil[1]{\orgdiv{State Key Laboratory of Precision Measuring Technology and Instruments}, \orgname{Tianjin University}, \city{Tianjin}, \country{China}}

\affil[2]{\orgdiv{Research Group Neuroinformatic}, \orgname{Faculty of Computer Science, University of Vienna}, \orgaddress{\city{Vienna}, \country{Austria}}}

\abstract{Motor imagery (MI) based EEG represents a frontier in enabling direct neural control of external devices and advancing neural rehabilitation. This study introduces a novel time embedding technique, termed traveling-wave based time embedding, utilized as a pseudo channel to enhance the decoding accuracy of MI-EEG signals across various neural network architectures. Unlike traditional neural network methods that fail to account for the temporal dynamics in MI-EEG in individual difference, our approach captures time-related changes for different participants based on a priori knowledge. Through extensive experimentation with multiple participants, we demonstrate that this method not only improves classification accuracy but also exhibits greater adaptability to individual differences compared to position encoding used in Transformer architecture. Significantly, our results reveal that traveling-wave based time embedding crucially enhances decoding accuracy, particularly for participants typically considered ``EEG-illiteracy". As a novel direction in EEG research, the traveling-wave based time embedding not only offers fresh insights for neural network decoding strategies but also expands new avenues for research into attention mechanisms in neuroscience and a deeper understanding of EEG signals. 
}


\maketitle

\section{Introduction}\label{sec1}

Brain-computer interfaces (BCIs) leveraging motor imagery (MI) enable the direct manipulation of external devices or computer applications through neural activities\cite{wolpaw2002brain}. Motor imagery involves the cognitive process of imagining specific movements, such as moving the left or right hand, without actual physical execution. Neuroscience research has established that, despite the absence of overt movement, the electroencephalogram (EEG) signals elicited during motor imagery closely resemble those observed during actual motor execution \cite{pfurtscheller2001motor}. Notably, significant event-related desynchronization (ERD) and synchronization (ERS) are observed in the motor cortex during MI, providing a robust foundation for decoding MI-EEG signals \cite{pfurtscheller2006mu}. Consequently, MI-based BCI systems facilitate motionless control of electronic devices, offering novel communication and interaction avenues for individuals with severe motor impairments \cite{hochberg2012reach}. Furthermore, MI-based BCIs hold considerable promise in rehabilitation medicine, particularly as innovative tools for neural rehabilitation to aid patients in recovering motor functions following strokes or brain injuries \cite{lazarou2018eeg}.

EEG, known for its high temporal resolution, is frequently employed to investigate the time-dependent changes in neural electrical signals following specific stimuli. For specific sensory, cognitive, or motor events, the elicited neural electrical signals are referred to as event-related potentials (ERPs), which are measured using EEG. In the case of specific visual stimuli, the corresponding ERPs, which are both time-locked and phase-locked, exhibit distinct ERP components such as P100, N100, P200, N200, and P300. For instance, in the field of cognitive neuroscience, EEG serves as a continuous, real-time physiological measure of cognitive load, capable of detecting subtle fluctuations in cognitive processing that traditional measures may overlook \cite{antonenko2010using}. However, the EEG signals associated with motor imagery are not phase-locked, making their ERP components less easily identifiable. Additionally, since the neural signals elicited by motor imagery tasks depend on the active participation of the individual and lack a feedback mechanism, it is challenging to pinpoint which segments of the EEG signals during the motor imagery period are most critical. 
For example, the duration of the imagery period is typically set between 3 to 5 seconds to balance the duration necessary for participants to generate discernible EEG patterns through endogenous cognitive activity \cite{cho2017eeg, kaya2018large,tangermann2012review}. 
However, individual differences such as attention span, reaction speed, and habitual motor imagery can lead to varying preferences for the duration of this imagery period among participants. Some studies \cite{pfurtscheller2001motor,ang2008filter} emphasize the necessity to account for these individual differences in EEG frequency bands and timing during the feature extraction phase, highlighting the diversity in EEG data across different moments within the imagery period.
In our previous research, we also found that moderately shortening the motor imagery task duration did not significantly reduce the accuracy of MI-EEG decoding \cite{miao2023weight}. 
However, the study did not identify which specific time segments of EEG data were most effective for decoding in different participants.

Recent advancements have underscored the pivotal role of artificial neural networks (ANNs) in decoding MI-EEG \cite{schirrmeister2017deep, lawhern2018eegnet}. While traditional manual feature construction methods have been foundational \cite{aflalo2015decoding}, techniques employing ANNs automate feature extraction and have significantly enhanced the accuracy of MI-EEG decoding \cite{miao2023lmda, miao2023weight}. Nevertheless, these neural network approaches often treat all EEG data within the imagery period as homogeneously significant, which may not reflect the underlying neural dynamics accurately.

Position encoding (or time embedding) might be a method worth considering. Although position encoding is a common technique in natural language processing \cite{vaswani2017attention, wu2021rethinking, shiv2019novel}, there is a notable absence of studies exploring the application of position encoding of EEG signals in motor imagery contexts. In natural language processing, positional encoding was introduced to address the issue of dependency between different parts of a sequence. Given the unique characteristics of EEG as a neuroimaging method, particularly its low signal-to-noise ratio, this study aims to explore whether time embedding can be utilized to identify the critical EEG segments that should be emphasized during decoding, especially in decoding methods dominated by artificial neural networks (ANN).
Investigating the role of time embedding in MI-EEG not only offers novel insights for enhancing MI-EEG decoding but also aids in a deeper understanding of the significant differences in EEG signals produced during various stages of motor imagery across different participants. This understanding could provide essential technical support for optimizing motor imagery EEG data collection and advancing the research into individual differences in MI-EEG signals.

To address this, we propose a novel traveling-wave based time embedding method, integrating it as a pseudo channel in EEG, to assess its impact on the decoding performance of convolutional neural networks for MI-EEG. For comparative analysis, we also employ position encoding method used in the Transformer \cite{vaswani2017attention} architecture. Our investigations span motor imagery datasets with a diverse range of channels, testing across different neural network architectures. Preliminary results suggest that the traveling-wave based time embedding method does not introduce extraneous noise into the MI-EEG signals and potentially enhances feature extraction by neural networks, provided the parameters are suitably chosen. Additionally, findings indicate that traveling-wave based time embedding may be more appropriate for MI-EEG data than the position encoding method employed in Transformer architectures. 
The time embedding approach proposed in this study holds promise for enhancing the decoding of EEG signals and for deepening our understanding of the underlying neuroscientific mechanisms of EEG activity.

\section{Methods}

\begin{figure*}
\centerline{\includegraphics[width=\textwidth]{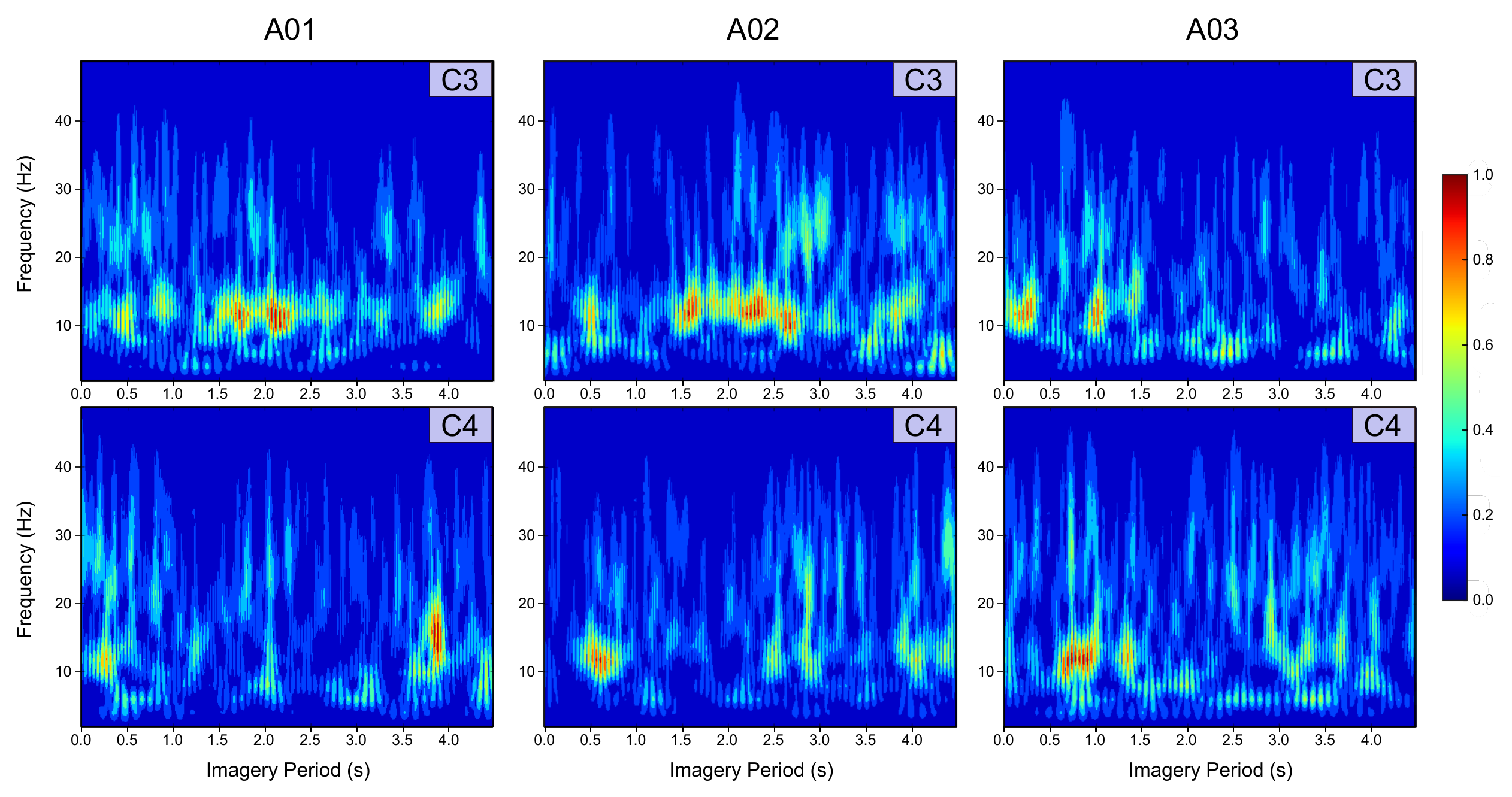}}
\caption{Normalized energy distribution of EEG signals across different time points and frequency bands during a single trial for various participants. The x-axis represents the time during the imagery period, with time zero marking the commencement of motor imagery. The energy in the time-frequency plots is normalized.}
\label{fig:time-frequency}
\end{figure*}

\subsection{Heuristic Development of time embedding}
In the analysis of MI-EEG signals, a common approach involves performing time-frequency analysis on EEG signals collected from different channels. By examining the differences in energy distribution across various frequency bands and time intervals, we can highlight the signal variations among different individuals during motor imagery. As shown in Figure \ref{fig:time-frequency}, the energy distribution in EEG signals varies across different time periods during the motor imagery task for different participants. The challenge in representing time embedding lies in the fact that the energy distribution differs across channels, making it difficult to incorporate the temporal distribution of EEG energy as prior information into an ANN. However, as depicted in Figure \ref{fig:time-frequency}, the temporal distribution of energy in MI-EEG resembles wave propagation, with slight differences in the location of peaks (high energy) across different participants or channels. This wave-like phenomenon can also be associated with attention during the motor imagery task. Since it is impossible to directly monitor the participant’s level of attention throughout the task, it is reasonable to hypothesize that some participants performed effective motor imagery in the early phase of the task, others in the middle phase, and some in the late phase. This bears a certain similarity to wave propagation.

\subsection{Traveling-wave based time embedding}

\begin{figure*}
\centerline{\includegraphics[width=\textwidth]{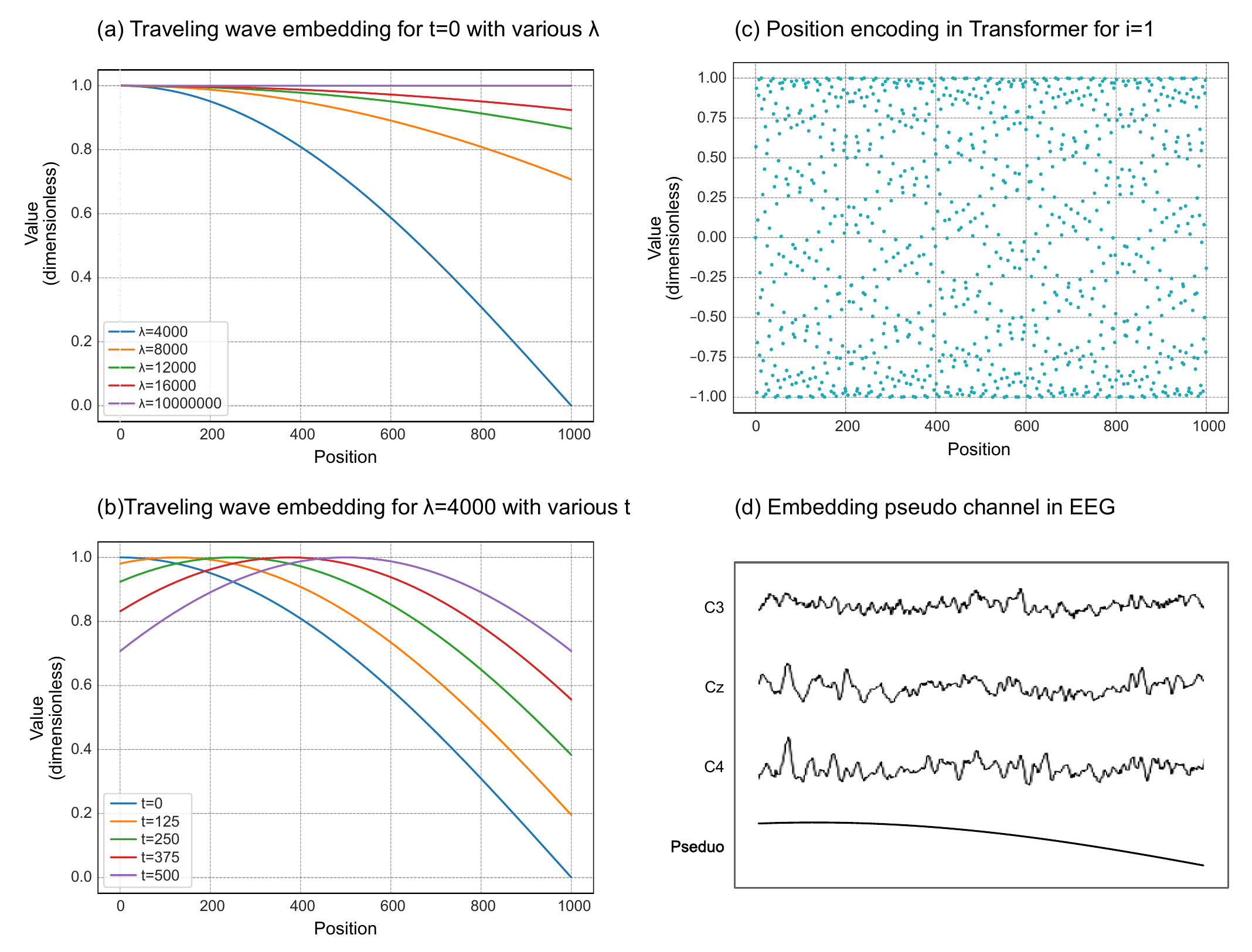}}
\caption{Visualization of Various Time-Embedding Results and Their Integration into EEG Signals. Subplot (a) illustrates the waveform variations of the traveling-wave based time embedding across different values of $\lambda$. Subplot (b) displays the waveform changes of the traveling-wave based time embedding with varying 
$t$ values. Subplot (c) uses a scatter plot to depict position encoding, highlighting its non-continuous distribution with each position uniquely encoded. Subplot (d) demonstrates the integration of time embedding as a pseudo channel within the EEG.}
\label{fig:pseudo}
\end{figure*}

The traveling-wave based time embedding is formulated as follows:
\begin{equation}
y = A \cos\left(\frac{2\pi (x - t)}{\lambda}\right)
\end{equation}
where \(A\) represents the amplitude, corresponding to the amplitude of the EEG signal, \(x\) denotes the position of EEG sampling points during the motor imagery period, \(t\) signifies the time offset, and \(\lambda\) indicates the wavelength. As illustrated in Figure \ref{fig:pseudo} panels (a) and (b), the choice of \(\lambda\) determines the rate of attenuation of the traveling-wave based time embedding within the imagery period—the larger the \(\lambda\), the slower the attenuation, and vice versa. The value of \(t\) determines the instance when the peak of the traveling-wave based time embedding occurs. This method of time embedding parallels the concept of a traveling wave equation, used here to simulate the varying temporal preferences of participants during the motor imagery period.

\subsection{Position encoding in Transformer}

Position encoding in the Transformer \cite{vaswani2017attention} is achieved by adding a position-dependent vector to the embedding representation of each input element. This allows the model not only to learn from the attributes of the input elements but also to consider their positions within the sequence. The position encoding in a Transformer is defined by the equations:
\begin{equation}
\text{PE}(pos, 2i) = \sin\left(\frac{pos}{10000^{2i/d}}\right)
\end{equation}
\begin{equation}
\text{PE}(pos, 2i+1) = \cos\left(\frac{pos}{10000^{2i/d}}\right)
\end{equation}
where \(pos\) is the position of the element within the sequence, \(i\) is the dimension index, and \(d\) is the dimensionality of the embedding. This encoding provides unique identifiers for different times within the imagery period by assigning distinct values to even and odd indices. Similarly to the traveling-wave based time embedding, position encoding in the Transformer is also integrated as a pseudo electrode in the preprocessed EEG data.

\subsection{Benchmark networks}

For this study, we selected two distinctly different artificial neural network architectures as benchmarks: Shallow-ConvNet \cite{schirrmeister2017deep}  (hereafter referred to as ConvNet)and EEGNet \cite{lawhern2018eegnet}. In ConvNet, the feature extraction network comprises a temporal convolution layer and a spatial convolution layer. In contrast, EEGNet includes an additional temporal convolution layer, with convolution layers that employ sparse connections. Furthermore, both ConvNet and EEGNet utilize kernel sizes in their spatial domain convolutions that match the number of EEG channels, effectively performing a full-connection operation. As indicated in the research by \cite{islam2020much}, such full-connection operations can learn the positional attributes of the input data, suggesting that the embedded pseudo electrodes can play a functional role during feature extraction and classification.

\section{Datasets}

\begin{figure*}
\centerline{\includegraphics[width=\textwidth]{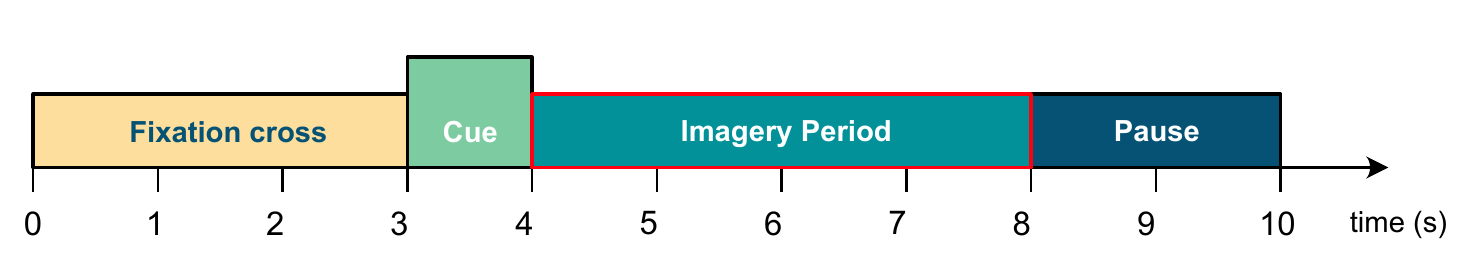}}
\caption{Typical EEG data collection process during motor imagery trials, illustrating the use of EEG signals acquired during the motor imagery period for decoding.}
\label{fig:scheme}
\end{figure*}

The protocol for collecting motor imagery EEG signals is highly standardized, involving sequential stages: fixation cross, cue signal, imagery period, and rest (as illustrated in Figure \ref{fig:scheme}). 
EEG data acquired in the imagery period are predominantly utilized for pattern recognition and classification tasks \cite{jeunet2016standard}. The time embedding methods proposed in the paper are also applied in imagery period. The EEG datasets used in this study also follow this protocol. 

\subsection{BCI4-2A Dataset}

In the BCI4-2A dataset\footnote{\url{www.bbci.de/competition/iv/\#dataset2a}}, EEG data were collected from nine healthy participants (ID A01-A09) using the international 10-20 system, with 22 EEG channels recorded at a sampling rate of 250 Hz. The data collection was performed using a BrainAmp MR plus amplifier and Ag/AgCl electrodes, and the signals were bandpass filtered between 0.5 Hz and 100 Hz, with a notch filter at 50 Hz to remove line noise.
Participants performed four motor imagery tasks: imagining movements of the left hand, right hand, both feet, and tongue. The data were collected in two separate sessions, each containing 288 trials (72 trials per task), with visual cues indicating the start of each motor imagery task. The data from the first session were designated for training, while the second session was reserved for testing.
The EEG data were labeled based on the type of motor imagery performed, with labels indicating the corresponding class (left hand, right hand, both feet, tongue) provided for each trial.

It is noteworthy that in the original ConvNet paper, the authors utilized EEG data from both the imagery period and the 0.5 seconds preceding it for decoding. However, our previous work \cite{miao2023weight} demonstrated that, with appropriate preprocessing methods, better performance can be achieved by using only the EEG data within the imagery period. This approach is therefore adopted in the current study.

\subsection{BCI4-2B Dataset}

In the BCI4-2B dataset\footnote{\url{www.bbci.de/competition/iv/\#dataset2b}}, EEG data were collected from nine healthy, right-handed participants (ID B01-B09) using three bipolar EEG electrodes (C3, Cz, and C4) at a sampling rate of 250 Hz. The recordings were made using a BrainAmp MR plus amplifier with a dynamic range of ±100 µV for the screening sessions and ±50 µV for the feedback sessions. The EEG signals were bandpass filtered between 0.5 Hz and 100 Hz, with a 50 Hz notch filter applied to remove line noise. 
Participants performed two motor imagery tasks: imagining movements of the left and right hands. Each session consisted of multiple runs, with each run comprising 20 trials per task, resulting in 120 trials per session. The data were labeled based on the motor imagery task being performed, with corresponding class labels (left hand or right hand) provided for each trial.
For the purposes of this study, the data from the first three sessions were used for training, while the last two sessions were reserved for testing, as described by \cite{miao2024time}. Only the EEG data recorded during the imagery period were extracted for decoding in this study.

\section{Experimental Settings}

\subsection{Data Preprocessing}

The preprocessing methods used in this paper continue from those employed in our previous work \cite{zhang2023priming} on motor imagery preprocessing. This involves a [4, 38] Hz 200-order Blackman window bandpass filter, data normalization, and Euclidean spatial alignment. Interested readers may refer to \cite{zhang2023priming, miao2023lmda} for more details.

\subsection{Experimental Environment and Parameters}

All experiments were conducted on a high-performance workstation equipped with an Intel Xeon(R) Gold 5117 CPU and an Nvidia Tesla V100 GPU, using the PyTorch framework. 
The training process employed minibatches of size 32. To ensure fair comparisons across different experimental methods, a unified random seed was used in all experiments to guarantee the repeatability of the experiments. Furthermore, all artificial neural networks were trained for 300 epochs, ensuring consistency.

Besides accuracy(\%), the kappa value are also employed as a key evaluation metric in this study to assess the performance of the models. 
\begin{equation}
    \kappa=\frac{acc-p_{0}}{1-p_{0}}
\end{equation}
where $acc$ is the accuracy of model testing, and $p_0$ is the random level accuracy. The kappa statistic provides a measure of agreement between the model’s predictions and the true labels, adjusting for the possibility of chance agreements. 

\section{Results}

The traveling-wave based time embedding technique utilizes two hyperparameters, \(t\) and \(\lambda\). In our experiments, the values for \(t\) were explored within the set \{0, 125, 250, 375, 500\}, and for \(\lambda\) within \{4000, 8000, 12000, 16000, 1000000\}. As participants exhibit different sensitivities to various hyperparameter combinations, Tables \ref{table:2a} and \ref{table:2b} respectively display the representative hyperparameter search results under two distinct neural network architectures. One set of results shows the best classification performance for different participants under various hyperparameter combinations, and the other illustrates the generalizability of a specific hyperparameter set within the traveling-wave based time embedding. The former set demonstrates the variation in participant performance following the integration of the traveling-wave based time embedding, while the latter reflects the method’s generalization across different participants. These tables also compare the outcomes against networks without any encoding and those utilizing classic position encoding from the Transformer architecture. To accurately reflect the impact of different hyperparameters in the traveling-wave based embedding on decoding accuracy, we further illustrate these differences using a violin plot in Figure \ref{fig:violin} and a heatmap in Figure \ref{fig:heat map}, which details the generalization of different hyperparameter combinations across participants.

\subsection{Results for BCI4-2A}

Table \ref{table:2a} presents the performance of two time encoding methods across two neural network architectures on the BCI4-2A dataset. It can be observed that for both the ConvNet and EEGNet architectures, traveling-wave based time embedding enhances the classification accuracy of the benchmark network to a certain extent. In contrast, position encoding in Transformer does not improve, and in some cases decreases, the classification performance. This suggests that convolutional neural networks can extract useful information from the time embedding, which also functions as a pseudo channel, and that the traveling-wave based time embedding is more suited to EEG signals than the position encoding method used in Transformer. Furthermore, time embedding strategies reveal differences among participants as well as between the neural network architectures. For instance, in the EEGNet architecture, traveling-wave based time embedding shows a significant advantage for participants A03 (12.5\% improvement) and A04 (16.7\% improvement). For the ConvNet architecture, the most notable improvement was observed in participant A06 (7.3\% improvement). Results from the Wilcoxon signed-rank test indicate significant differences between the ConvNet and EEGNet architectures. In the BCI4-2A dataset, ConvNet exhibited significant performance improvements with the traveling-wave based time embedding strategy (P$<$0.05), whereas EEGNet did not show significant differences (P$>$0.05).

\begin{table*}[]
\caption{Binary classification performance (\% for accuracy) of different algorithms on BCI4-2A}
\centering
\label{table:2a}
\resizebox{1.0\textwidth}{!}{ %
\begin{threeparttable}
\begin{tabularx}{1.4\textwidth}{lcccccccccccc} %
\hline
\textbf{Methods} & \textbf{A01} & \textbf{A02} & \textbf{A03} & \textbf{A04} & \textbf{A05} & \textbf{A06} & \textbf{A07} & \textbf{A08} & \textbf{A09} & \textbf{mean(kappa)$\pm$std} & \textbf{P-value} \\ \hline
C\_Benchmark                   & 89.2 & 66.3 & 94.1 & 83.7 & 69.1 & 58.7 & 96.5 & 91.3 & 87.5 & 81.8(0.757)$\pm$13.6         & 0.017 \\
C\_Transformer       & 88.9 & 66.3 & 92.4 & 83.3 & 70.1 & 59.7 & 96.5 & 88.5 & 88.2 & 81.6(0.755)$\pm$12.9         & 0.012 \\
C\_t375\_$\lambda$16000                & 89.6 & 66.7 & 95.5 & 85.8 & 63.9 & 65.6 & 96.2 & 89.2 & 87.2 & 82.2(0.763)$\pm$13.1         & 0.012 \\
C\_TravelingWave      & 92.4 & 68.1 & 95.5 & 87.5 & 71.5 & 66.0 & 96.5 & 91.0 & 91.0 & 84.4(0.792)$\pm$12.3         & -     \\ \hline
E\_Benchmark                    & 79.5 & 68.8 & 83.3 & 52.4 & 65.3 & 60.1 & 88.6 & 81.3 & 86.8 & 74.0(0.653)$\pm$12.8         & 0.129 \\
E\_Transformer         & 75.0 & 60.4 & 95.5 & 56.9 & 58.0 & 57.3 & 83.0 & 82.3 & 85.4 & 72.6(0.635)$\pm$14.7         & 0.004 \\
E\_t125\_$\lambda$16000                 & 79.9 & 63.5 & 93.4 & 65.3 & 58.3 & 56.9 & 87.5 & 80.2 & 87.2 & 74.7(0.663)$\pm$13.8         & 0.004 \\ 
E\_TravelingWave      & 82.3 & 68.1 & 95.8 & 69.1 & 61.8 & 59.0 & 92.4 & 83.7 & 87.9 & 77.8(0.704)$\pm$13.6         & -     \\ \hline
\end{tabularx}
\begin{tablenotes}
\small
\item ``P-value" indicates the Wilcoxon signed-rank test results, comparing the best performing traveling-wave based time embedding configuration to others within the same ANN.
\item In the ``Methods" column, `C' denotes the abbreviation for ConvNet, while `E' represents the abbreviation for EEGNet.
\item ``\_Benchmark" refers to ANNs that do not include any form of time embedding, serving as control models.
\item ``\_Transformer" represents ANNs that incorporate classical position encoding, as utilized in Transformer models.
\item ``\_t375\_$\lambda$16000" identifies a specific hyperparameter set in the traveling-wave based time embedding that demonstrated the most effective generalization across participants.
\item ``\_TravelingWave" marks the configuration under the traveling-wave based time embedding that yielded the highest performance for the participant.
\end{tablenotes}
\end{threeparttable}
}
\end{table*}

\subsection{Results for BCI4-2B}

The BCI4-2B dataset employed the same analysis methods as BCI4-2A. From Table \ref{table:2b}, traveling-wave based time embedding also proved to be beneficial. For example, the average classification accuracy of ConvNet with the hyperparameters \(t=125\), \(\lambda=4000\) is 82.6\%, while the average classification accuracy of ConvNet without time embedding is 81.8\%. For the EEGNet network, the optimal hyperparameter combination \(t=250\), \(\lambda=4000\) yielded an average classification accuracy of 85.5\% in BCI4-2B. Similarly, compared to position encoding in Transformer, traveling-wave embedding generally showed better classification performance for most participants in BCI4-2B. The time embedding strategy also demonstrated variations among participants and differences in neural network architecture responses. Notably, although the benchmark architectures of ConvNet and EEGNet performed differently compared to BCI4-2A, their responses to the time embedding strategy also varied from BCI4-2A. Results from the Wilcoxon signed-rank test reveal that ConvNet did not exhibit significant changes with the travel-wave based time embedding strategy (P$>$0.05), whereas EEGNet showed significant changes (P$<$0.05). Regardless of the architecture, whether ConvNet or EEGNet, the classification performance of participants B02 and B03 was significantly enhanced. Compared to other participants, B02 and B03 are often referred to as ``EEG-illiteracy" because their EEG signals are typically harder to decode.

\begin{table*}[]
\caption{Binary classification performance (\% for accuracy) of different algorithms on BCI4-2B}
\centering
\label{table:2b}
\resizebox{\textwidth}{!}{ %
\begin{threeparttable}
\begin{tabularx}{1.4\textwidth}{lcccccccccccc} %
\hline
\textbf{Methods} & \textbf{B01} & \textbf{B02} & \textbf{B03} & \textbf{B04} & \textbf{B05} & \textbf{B06} & \textbf{B07} & \textbf{B08} & \textbf{B09} & \textbf{mean(kappa)$\pm$std} & \textbf{P-value} \\ \hline
C\_Benchmark     & 80.9                    & 56.1                    & 59.7                    & 97.5                    & 90.3                    & 89.4                    & 82.2                    & 92.2                    & 87.5                    & 81.8(0.636)$\pm$14.4                    & 0.359                       \\
C\_Transformer   & 76.9                    & 61.4                    & 66.9                    & 97.5                    & 91.6                    & 87.5                    & 82.8                    & 91.3                    & 87.5                    & 82.6(0.652)$\pm$12.0                    & 0.129                       \\
C\_t125\_$\lambda$4000   & 79.7                    & 62.1                    & 70.0                    & 96.9                    & 92.5                    & 83.1                    & 84.4                    & 90.3                    & 84.7                    & 82.6(0.652)$\pm$10.9                    & 0.017                       \\
C\_TravelingWave & 79.7                    & 63.6                    & 72.5                    & 97.8                    & 92.5                    & 86.3                    & 85.6                    & 91.9                    & 86.3                    & 84.0(0.680)$\pm$10.6                    & -                           \\ \hline
E\_Benchmark      & 80.0                    & 60.0                    & 70.6                    & 97.8                    & 93.8                    & 90.0                    & 86.6                    & 92.8                    & 91.9                    & 84.8(0.696)$\pm$12.4                    & 0.012                       \\
E\_Transformer    & 81.9                    & 62.1                    & 69.4                    & 97.5                    & 92.5                    & 85.3                    & 82.8                    & 93.1                    & 91.9                    & 84.1(0.682)$\pm$11.7                    & 0.012                       \\
E\_t250\_$\lambda$4000    & 80.9                    & 64.6                    & 74.4                    & 97.8                    & 94.1                    & 87.2                    & 85.9                    & 94.1                    & 90.3                    & 85.5(0.710)$\pm$10.6                    & 0.004                       \\
E\_TravelingWave  & 83.1                    & 68.2                    & 75.3                    & 98.1                    & 95.3                    & 99.1                    & 87.5                    & 95.3                    & 91.3                    & 88.1(0.762)$\pm$10.7                    & -                          
   \\ \hline
\end{tabularx}
    \begin{tablenotes}
        \small
        \item ``P-value" indicates the Wilcoxon signed-rank test results, comparing the best performing traveling-wave based time embedding configuration to others within the same neural network architecture.
        \item In the ``Methods" column, `C' denotes the abbreviation for ConvNet, while `E' represents the abbreviation for EEGNet.
        \item ``\_Benchmark" refers to ANNs that do not include any form of time embedding, serving as control models.
        \item  ``\_Transformer" represents ANNs that incorporate classical position encoding, as utilized in Transformer models.
        \item ``\_t125\_$\lambda$4000" identifies a specific hyperparameter set in the traveling-wave based time embedding that demonstrated the most effective generalization across participants.
        \item ``\_TravelingWave" marks the configuration under the traveling-wave based time embedding that yielded the highest performance for the participant.
    \end{tablenotes}
\end{threeparttable}
}
\end{table*}

\subsection{Impact of hyperparameter variations on participant decoding accuracy}

\begin{figure*}[h]
\centerline{\includegraphics[width=\textwidth]{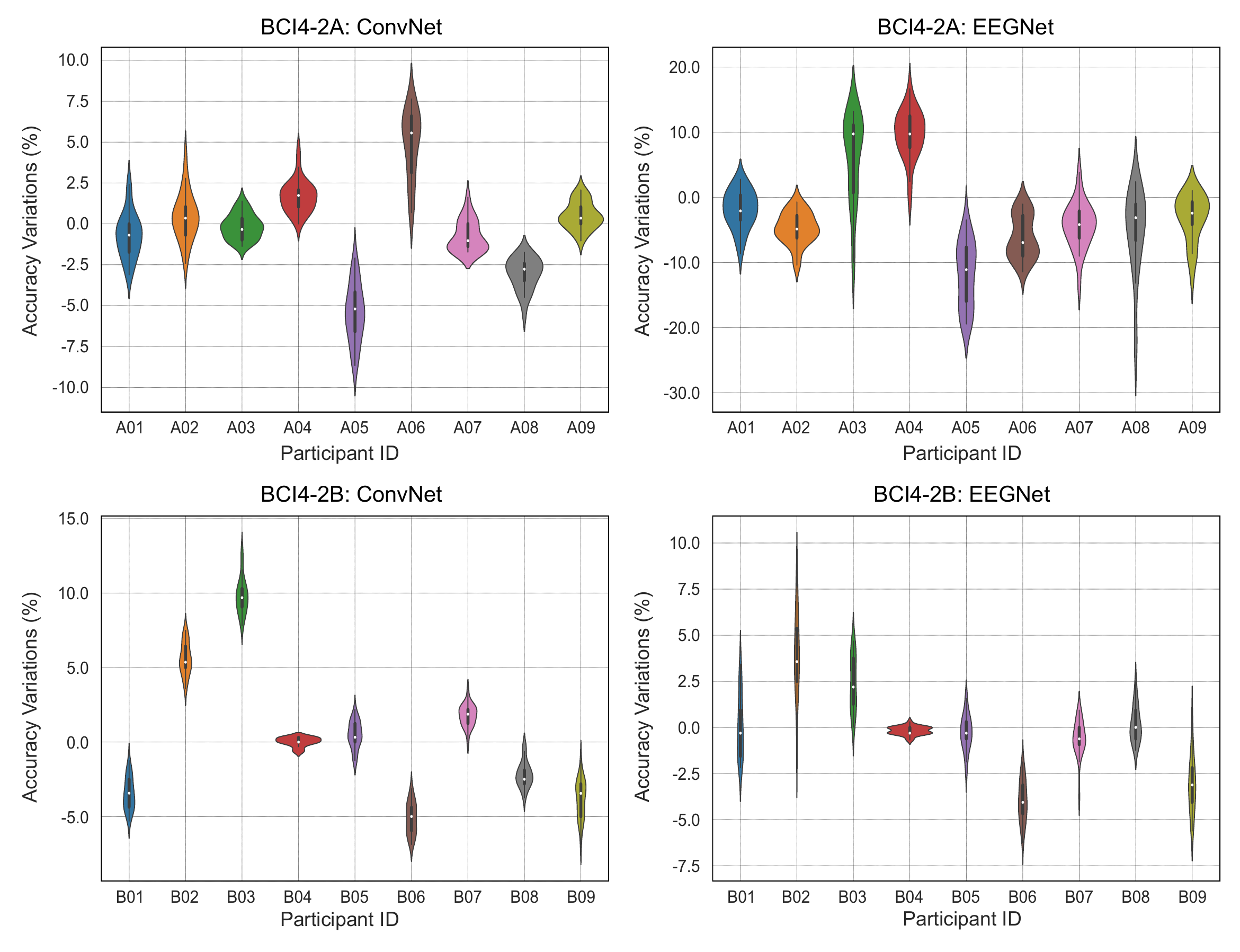}}
\caption{Variations in decoding accuracy across participants under different hyperparameters of traveling-wave based time embedding compared to benchmark networks. Positive values on the vertical axis indicate higher decoding accuracy compared to the benchmark network, whereas negative values signify lower decoding accuracy than the benchmark network.}
\label{fig:violin}
\end{figure*}

To comprehensively examine the effects of hyperparameter values in traveling-wave based time embedding on decoding accuracy across different participants, Figure \ref{fig:violin} presents violin plots that illustrate the variability in decoding accuracy with different hyperparameter combinations compared to a benchmark network architecture. The values for \(t\) in the traveling-wave based time embedding were explored within the set \{0, 125, 250, 375, 500\}, and for \(\lambda\) within \{4000, 8000, 12000, 16000, 1000000\}, resulting in 25 experimental outcomes for each participant across each network architecture. Figure \ref{fig:violin} vividly captures the variations in decoding accuracy post-time embedding, as well as the differential impacts of various decoding networks and electrode datasets on the performance of time embedding.

A ``tall and slim" violin plot indicates that a participant is particularly sensitive to the hyperparameter values, whereas a ``short and wide" violin plot suggests a lower sensitivity. Figure \ref{fig:violin} also reveals that some participants, such as B02 and B03, experience significant improvements in decoding performance following the application of time embedding. However, not all participants benefit from the time embedding strategy; for example, participants A05 and B06 show a decrease in decoding performance after the application of time embedding. The majority of participants' decoding outcomes depend on the choice of hyperparameters in the time embedding, and selecting appropriate hyperparameter combinations can substantially enhance decoding performance for these individuals.

\subsection{Hyperparameter analysis in traveling-wave based time embedding}

\begin{figure*}[h]
\centerline{\includegraphics[width=\textwidth]{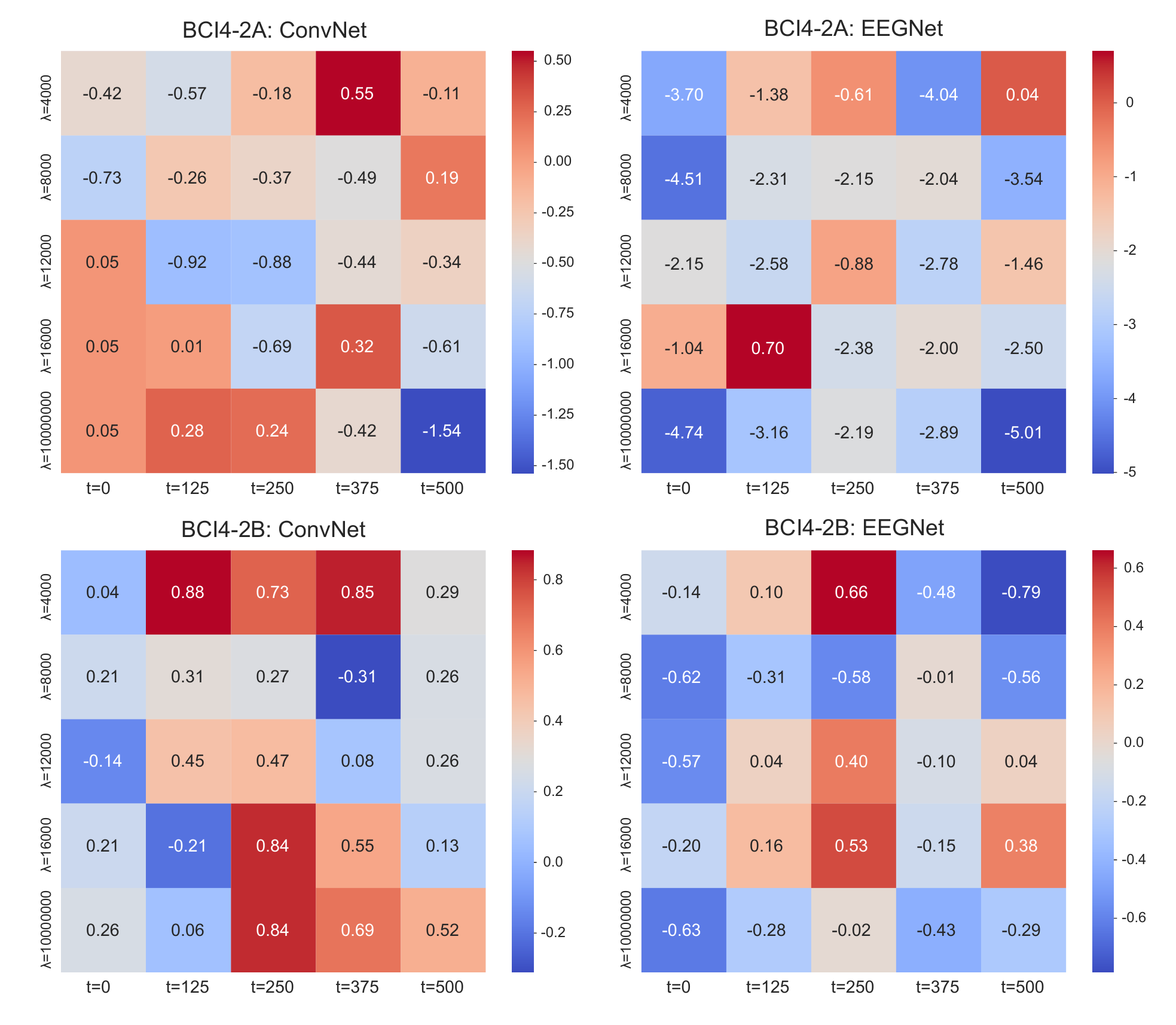}}
\caption{Average accuracy variations (across all participants) by parameter combinations in traveling-wave based time embedding.Positive values on the vertical axis indicate higher decoding accuracy compared to the benchmark network, whereas negative values signify lower decoding accuracy than the benchmark network.}
\label{fig:heat map}
\end{figure*}

As depicted in Figure \ref{fig:violin}, the application of traveling-wave based time embedding resulted in improved decoding accuracy for some participants, while others experienced a decrease. Consequently, Figure \ref{fig:heat map} examines the generalizability of traveling-wave based time embedding across different network architectures. The heatmap in Figure \ref{fig:heat map} uses a zero baseline, representing the average accuracy across all participants for the benchmark within the dataset. This visualization demonstrates that the ConvNet architecture exhibits greater robustness compared to EEGNet, evidenced by smaller fluctuations in average accuracy across the dataset under varying hyperparameter settings of traveling-wave based time embedding. 

On the other hand, compared to the results in Tables \ref{table:2a} and \ref{table:2b}, the overall increase in average accuracy across the dataset is modest, suggesting that different participants have distinct preferences for the hyperparameters in time embedding. This indirectly supports the variability in temporal preferences among participants during motor imagery. For ConvNet, a robust hyperparameter combination is \(\lambda = 4000\) and \(t = 375\), which consistently improved average accuracy across two different datasets. In contrast, for EEGNet, an increase in \(\lambda\) to 1000000 generally led to a decline in average accuracy across various combinations, indicating limited recognition capabilities for position encoding. This may relate to EEGNet’s network architecture, which features sparse connections and a spatial convolution followed by an additional temporal convolution layer.

\section{Conclusion}\label{sec13}
This study has examined the efficacy of time embedding methods in the context of MI-EEG decoding, with a particular focus on a novel traveling-wave based time embedding approach integrated as a pseudo channel into MI-EEG signals. Extensive validations were carried out in two influential artificial neural network architectures and two datasets with significant variations in the number of channels. The experimental results confirm that the traveling-wave based time embedding method significantly enhances decoding accuracy, especially when optimal hyperparameters are selected. This improvement is especially pronounced in specific subsets of participants, especially for some participants who are considered as ``EEG-illiteracy". Moreover, this method outperforms the position encoding method used in Transformer architecture in terms of performance and robustness across diverse network architectures and motor imagery datasets. 

However, the traveling-wave based time embedding model proposed in this study is a single-peak model, and only a limited 5×5 hyperparameter search was conducted for the parameters. The effectiveness of time embedding in MI-EEG was solely evaluated based on decoding accuracy, which significantly limits its application in enhancing the interpretability of EEG data. Future research will focus on two main directions: (1) exploring the underlying neural mechanisms of the optimal time embedding model from perspectives such as time-frequency analysis and participant attention mechanisms; (2) investigating the impact of time embedding, when integrated as a pseudo-channel into EEG signals, on the feature extraction process of artificial neural networks, with an emphasis on model interpretability.

\section*{Acknowledgements}
This work was supported in part by Ernst-Mach scholarship. 
We extend our gratitude to Yiwei Yang for polishing the figures. 

\section*{Declarations}
\subsubsection*{Conflict of interest}
The authors declare that they have no known competing financial interests or personal relationships that could have appeared to influence the work reported in this paper.




\bibliography{sn-bibliography}

\end{document}